\documentclass[aps,pre,superscriptaddress,unsortedaddress,twocolumn,showpacs]{revtex4}

\usepackage{graphicx,color,mathrsfs}
\begin{document}
\def\be{\begin{equation}}
\def\ee{\end{equation}}
\def\bea{\begin{eqnarray}}
\def\eea{\end{eqnarray}}
\def\fr{\frac}
\def\l{\label}
\renewcommand*{\thefootnote}{\arabic{footnote}}

%%%%%%%%%%%%%%%%%%%%%%%%%%%%%%%%%%%%%%%%%%%%%%%%%%%%%%%%%%%%%%%%%%%%%%%%%%%%%%%%%%%%%%%%%
\title{Path-integral formalism for stochastic resetting:\\ Exactly
solved examples and shortcuts to confinement}
\author{\'{E}dgar Rold\'{a}n\footnote{Corresponding author:
edgar@pks.mpg.de}}
\affiliation{Max-Planck Institute for the Physics of Complex Systems, cfAED and GISC, N\"{o}thnitzer Stra\ss e 38, 01187 Dresden, Germany}
\author{Shamik Gupta}
\affiliation{Department of Physics, Ramakrishna Mission Vivekananda
University, Belur Math, Howrah 711 202, West Bengal, India}
\begin{abstract}
We study the dynamics of overdamped Brownian particles diffusing in
conservative force fields and undergoing stochastic resetting to a given
location with a generic space-dependent rate of resetting. We present a
systematic approach involving path integrals and elements
of renewal theory that allows to derive analytical expressions for a variety of statistics of the dynamics
such as (i) the propagator prior to first reset; (ii) the distribution
of the first-reset time, and (iii) the spatial distribution of the
particle at long times. We apply our approach to
several representative and hitherto unexplored examples of resetting
dynamics. A particularly interesting example for which we find
analytical expressions for the statistics of resetting is that of a Brownian particle trapped in a harmonic potential with a rate of resetting that depends on the instantaneous energy of the particle.
We find that using energy-dependent resetting processes is more
effective in achieving spatial confinement of Brownian particles on a
faster timescale than by performing quenches of parameters of the harmonic potential.
\end{abstract}
\date{\today}
\maketitle

%%%%%%%%%%%%%%%%%%%%%%%%%%%%%%%%%%%%%%%%%%%%%%%%%%%%%%%%%%%%%%%%%%%%%%%%%%%%%%%%
\section{Introduction}
\l{sec:intro}
Changes are inevitable in nature, and those that are most dramatic with often 
drastic consequences are the ones that occur {\em all of a sudden}. A
particular class of such changes comprises those in which the system
during its temporal evolution makes a sudden jump (a ``reset") to a
fixed state or configuration. Many nonequilibrium processes are
encountered across disciplines, e.g., in physics, biology, and
information processing, which involve sudden transitions between different states or configurations. 
The erasure of a bit of information~\cite{Landauer:1961,Bennett:1973} by
mesoscopic machines may be thought of as a physical process in which a
memory device that is strongly affected by thermal fluctuations resets
its state (0 or 1) to a prescribed erasure
state~\cite{Berut:2012,Mandal:2012,Roldan:2014,Koski:2014,Fuchs:2016}.
In biology, resetting plays an important role inter alia in sensing of
extracellular ligands by single cells~\cite{Mora:2015}, and in
transcription of genetic information by macromolecular enzymes called
RNA polymerases~\cite{Roldan:2016}. During RNA transcription, the
recovery of RNA polymerases from inactive transcriptional pauses is a
result of a kinetic competition between diffusion and resetting of the
polymerase to an active state via RNA cleavage~\cite{Roldan:2016}, as
has been recently tested in high-resolution single-molecule
experiments~\cite{Lisica:2016}. Also, there are ample examples of
biochemical processes that initiate (i.e., reset) at random so-called {\em stopping}
times~\cite{Gillespie:2014,Hanggi:1990,Neri:2017}, with the initiation
at each instance occurring in different regions of space ~\cite{Julicher:1997}. In addition,
interactions play a key role in determining when and where a chemical
reaction occurs~\cite{Gillespie:2014}, a fact that affects the
statistics of the resetting process. For instance, in the above
mentioned example of recovery of RNA polymerase by the process of
resetting, the interaction of the hybrid DNA-RNA may alter the time that a polymerase takes to recover from its
inactive state~\cite{Zamft:2012}. It is therefore quite pertinent and
timely to study resetting of mesoscopic systems that evolve under the influence
of external or conservative force fields.

Simple diffusion subject to resetting to a given location at random
times has emerged in recent
years as a convenient theoretical framework to discuss the phenomenon of
stochastic resetting
\cite{Evans:2011-1,Evans:2011-2,Evans:2014,Christou:2015,Eule:2016,Nagar:2016}.
The framework has later been generalized to consider different choices
of the resetting position \cite{Boyer:2014,Majumdar:2015-2}, resetting of continuous-time random walks
 \cite{Montero:2013,Mendez:2016}, L\'{e}vy \cite{Kusmierz:2014} and
exponential constant-speed flights \cite{Campos:2015}, time-dependent
resetting of a Brownian particle \cite{Pal:2016}, and in
discussing memory effects \cite{Boyer:2017} and phase transitions in
reset processes \cite{Harris:2017}. Stochastic resetting has also been invoked in the context of many-body
dynamics, e.g., in reaction-diffusion models \cite{Durang:2014}, fluctuating
interfaces \cite{Gupta:2014,Gupta:2016}, interacting Brownian motion
\cite{Falcao:2017}, and in discussing optimal search times in a crowded environment \cite{Kusmierz:2015,
Reuveni:2016,Bhat:2016,Pal:2017}.
However, little is known about the statistics of stochastic resetting of
Brownian particles that diffuse under the influence of force
fields~\cite{Pal:2015}, and that too
in presence of a rate of resetting that varies in space.

\begin{figure*}[!ht]
\centering
\includegraphics[width=0.9\textwidth]{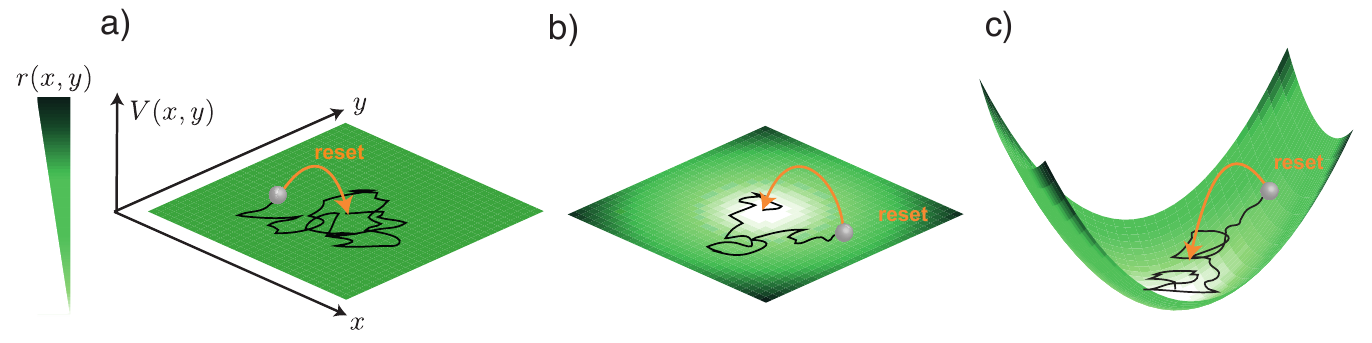} \caption{{\bf
Stochastic resetting of Brownian particles moving in force fields and
subject to stochastic resetting with a rate that may depend
on space}. Illustration of the motion of a
single overdamped Brownian particle (grey sphere) that diffuses in two dimensions $(x,y)$ in presence of a conservative potential $V(x,y)$.
The particle traces a stochastic trajectory (black curve) in the
two-dimensional space until its position is reset at a random instant of
time to its
initial value (orange arrow); subsequent to the reset, the particle resumes its
stochastic motion until the next reset happens. The rate of resetting
$r(x,y)$ (left colorbar) may depend on the position of the
particle. We study three  different scenarios: \textbf{a)} Resetting of
free Brownian particles under a space-independent rate of resetting;
\textbf{b)} Resetting of free Brownian particles under a space-dependent
rate of resetting; \textbf{c)} Resetting of Brownian particles moving in force fields under a space-dependent rate of resetting.}
\l{fig:fig1}
\end{figure*}

In this paper, we study the dynamics of overdamped Brownian particles
immersed in a thermal environment, which diffuse under the influence of a force field, and whose position may be
stochastically reset to a given spatial location with a rate of
resetting that has an essential dependence on space. We use an approach that allows to obtain exact expressions for the transition probability prior to the first
reset, the first reset-time distribution, and, most importantly, the
stationary spatial distribution of the particle. The approach is based
on a combination of the theory of renewals \cite{Cox:1962} and the
Feynman-Kac path-integral formalism of treating stochastic processes
\cite{Feynman:2010,Schulman:1981,Kac:1949,Kac:1951}, and
consists in a mapping of the dynamics of the Brownian resetting problem
to a suitable quantum mechanical evolution in imaginary time. We note
that the Feynman-Kac formalism has been applied extensively in the past
to discuss dynamical processes involving diffusion \cite{satya}, and has to the best
of our knowledge not been applied to discuss stochastic resetting. To
demonstrate the utility of the approach, we consider several different
stochastic resetting problems, see Fig. \ref{fig:fig1}: i) Free Brownian particles
subject to a space-independent rate of resetting (Fig.
\ref{fig:fig1}a)); ii) Free Brownian particles subject to
resetting with a rate that depends quadratically on the distance to the
origin (Fig. \ref{fig:fig1}b)); and iii) Brownian particles
trapped in a harmonic potential and undergoing reset events with a rate
that depends on the energy of the particle (Fig.
\ref{fig:fig1}c)). In this paper, we consider for purposes of
illustration the corresponding scenarios in one dimension, although our
general approach may be extended to higher dimensions. Remarkably, we
obtain exact analytical expressions in all cases, and, notably, in cases ii) and iii), where a
standard treatment of analytic solution by using the Fokker-Planck
approach may appear daunting, and whose relevance in physics may be
explored in the context of, e.g., optically-trapped colloidal particles
and hopping processes in glasses and gels. We further explore the
dynamical properties of case iii), and compare the relaxation properties
of dynamics corresponding to potential energy quenches and due to sudden activation of space-dependent stochastic resetting.

%%%%%%%%%%%%%%%%%%%%%%%%%%%%%%%%%%%%%%%%%%%%%%%%%%%%%%%%%%%%%%%%%%%%%%%%%%%%%%%%
\section{General formalism}
\l{sec:general-formalism}

%%%%%%%%%%%%%%%%%%%%%%%%%%%%%%%%%%%%%%%%%%%%%%%%%%%%%%%%%%%%%%%%%%%%%%%%%%%%%%%%
%%%%%%%%%%%%%%%%%%%%%%%%%%%%%%%%%%%%%%%%%%%%%%%%%%%%%%%%%%%%%%%%%%%%%%%%%%%%%%%%
\subsection{Model of study: resetting of Brownian particles diffusing in force fields}
\l{sec:model}

Consider an overdamped Brownian particle diffusing in one dimension $x$ in presence of
a time-independent force field $F(x)=-\partial_x V(x)$, with $V(x)$
denoting a potential energy landscape.  The dynamics of the particle is described by a Langevin equation of the form 
\be
\frac{{\rm d}x}{{\rm d}t}=\mu F(x)+\eta(t),
\l{eq:eom}
\ee
where $\mu$ is the mobility of the particle, defined as the
velocity per unit force. In Eq.~(\ref{eq:eom}), $\eta(t)$ is a
Gaussian white noise, with the properties
\be
\langle\eta(t)\rangle=0,~\langle\eta(t)\eta(t')\rangle=2D\delta(t-t'),
\ee
where $\langle\, \cdot\, \rangle$ denotes average over noise
realizations, and $D\! >\! 0$ is the diffusion coefficient of the
particle, with the dimension of length-squared over time. We assume that
the Einstein relation holds: $D=k_{\rm B}T \mu$, with $T$ being the
temperature of the environment, and with $k_{\rm B}$ being the Boltzmann constant. In addition to the dynamics (\ref{eq:eom}), the particle is subject to a stochastic resetting dynamics with a space-dependent
 resetting rate $r(x)$, whereby, while at position $x$ at time $t$, the
particle in the ensuing infinitesimal time interval ${\rm d}t$ either
follows the dynamics (\ref{eq:eom}) with probability $1-r(x){\rm d}t$, or resets to a given reset destination $x^{(\rm r)}$ with
probability $r(x){\rm d}t$. Our analysis holds for any arbitrary reset function
$r(x)$, with the only obvious constraint $r(x) \ge 0~\forall\,x$;
moreover, the formalism may be generalized to higher
dimensions. In the following, we consider the reset location to be the same
as the initial location $x_0$ of the particle, that is, $x^{(\rm r)}=x_0$.

A quantity of obvious interest and relevance is the spatial distribution
of the particle: What is the probability $P(x,t|x_0,0)$ that the particle is at
position $x$ at time $t$, given that its initial location is $x_0$?
From the dynamics given in the preceding paragraph, it is straightforward
to write the time evolution equation of $P(x,t|x_0,0)$:
\bea
&&\hspace{-0.5cm}\frac{\partial P(x,t|x_0,0)}{\partial t}=-\mu \frac{\partial
(F(x)P(x,t|x_0,0))}{\partial x}+D\frac{\partial^2
P(x,t|x_0,0)}{\partial x^2}\nonumber\\
&&\hspace{-0.5cm}-r(x)P(x,t|x_0,0)+\int {\rm
d}y~r(y)P(y,t|x_0,0)\delta(x-x_0),
\l{eq:timevolution}
\eea
where the first two terms on the right hand side account for the
contribution from the diffusion of the particle in the force field
$F(x)$, while the last two terms stand for the contribution owing to the resetting of the particle:
the third term represents the loss in probability arising from the resetting of
the particle to $x_0$, while the fourth term denotes the gain in
probability at the location $x_0$ owing to resetting from all locations $x
\ne x_0$. When exists, the stationary distribution $P_{\rm st}(x|x_0)$
satisfies
\bea
&& 0=-\mu \frac{\partial
(F(x)P_{\rm st}(x|x_0))}{\partial x}+D\frac{\partial^2
P_{\rm st}(x|x_0)}{\partial x^2}\nonumber\\
&& -r(x)P_{\rm st}(x|x_0)+\int {\rm
d}y~r(y)P_{\rm st}(y|x_0)\delta(x-x_0).
\l{eq:stationary}
\eea
It is evident that solving for either the time-dependent distribution $P(x,t|x_0,0)$ or the
stationary distribution $P_{\rm st}(x|x_0)$ from Eqs.
(\ref{eq:timevolution}) and (\ref{eq:stationary}), respectively, is a
formidable task even with $F=0$, unless the function $r(x)$ has simple forms. For
example, in Ref.~\cite{Evans:2011-2}, the authors considered a 
solvable example with $F(x)=0$, where the function $r(x)$ is zero in a window around $x_0$ and is
constant outside the window.  

In this work, we employ a different approach to solve for the stationary spatial distribution, by 
invoking the path integral formalism of quantum mechanics and by using
elements of the theory of renewals. In
this approach, we compute $P_{\rm st}(x|x_0)$, the stationary
distribution {\em in presence of reset events}, in terms of suitably-defined
functions that take into account the occurrence of trajectories that
evolve {\em without undergoing any reset events} in a
given time, see Eq. (\ref{eq:Pxstat-final})
below. This approach provides a viable alternative to obtaining the
stationary spatial distribution by solving the
Fokker-Planck equation (\ref{eq:stationary}) that explicitly takes into
account the occurrence of trajectories that
evolve {\em while undergoing reset events} in a
given time. As we will
demonstrate below, the method allows to obtain exact expressions even in
cases with nontrivial forms of $F(x)$ and $r(x)$.

%%%%%%%%%%%%%%%%%%%%%%%%%%%%%%%%%%%%%%%%%%%%%%%%%%%%%%%%%%%%%%%%%%%%%%%%%%%%%%%%
%%%%%%%%%%%%%%%%%%%%%%%%%%%%%%%%%%%%%%%%%%%%%%%%%%%%%%%%%%%%%%%%%%%%%%%%%%%%%%%%
\subsection{Path-integral approach to stochastic resetting}
\l{sec:quantities-of-interest}

Here, we invoke the well-established path-integral approach based on the Feynman-Kac formalism to
discuss stochastic resetting. To proceed, let us first consider a representation of the dynamics in discrete times $t_i=i\Delta t$,
with $i=0,1,2,\ldots$, and $\Delta t>0$ being a small time step. The
dynamics in discrete times involves the particle at position $x_i$ at
time $t_i$ to either reset and be at $x^{(\rm r)}$ at the next time
step $t_{i+1}$ with probability $r(x_i)\Delta t$ or follow the dynamics
given by Eq.~(\ref{eq:eom})
with probability $1-r(x_i)\Delta t$. The position 
of the particle at time $t_i$ is thus given by
\bea
x_i & =
\left\{
\begin{array}{ll}
x_{i-1}+\Delta t\left(\mu \bar{F}(x_i)+\eta_i\right)& {\rm
with\,\, prob.}\;1-r(x_{i-1})\Delta t,\\
x^{({\rm r})} & {\rm with\,\, prob.}\; r(x_{i-1})\Delta t,\\
               \end{array}
        \right. 
\l{eq:dynamics}        
\eea
where we have defined $\bar{F}(x_i)\equiv (F(x_{i-1})+F(x_i))/2$, and have used the Stratonovich rule in discretizing the
dynamics (\ref{eq:eom}), and where the time-discretized Gaussian, white noise $\eta_i$
satisfies 
\be
\langle\eta_i\eta_j\rangle=\sigma^2\delta_{ij},
\ee
with $\sigma^2$ a positive constant with the dimension of length-squared over time-squared. In particular, the joint probability distribution
of occurrence of a given realization $\{\eta_i\}_{1\le i\le N}$ of the
noise, with $N$ being a positive integer, is given by 
\be
P[\{\eta_i\}]=\left(\frac{1}{2\pi
\sigma^2}\right)^{N/2}\exp\left(-\frac{1}{2\sigma^2}\sum_{i=1}^{N}\eta_i^{2}\right).
\l{eq:joint-distribution}
\ee
In the absence of any resetting and forces, the displacement of the particle at time $t\equiv N\Delta t$
from the initial location is given by $\Delta x \equiv x_N-x_0=\Delta t\sum_{i=1}^{N}\eta_i$, 
so that the mean-squared displacement is
$\langle(\Delta x)^{2}\rangle=\sigma^2 N (\Delta t)^2$.
In the continuous-time limit, $N\to\infty,\Delta t\to0$, keeping the
product $N\Delta t$ fixed and finite and equal to $t$, the mean-squared displacement becomes
$\langle(\Delta x)^{2}\rangle=2Dt$, with $D\equiv \lim_{\sigma \to
\infty,\Delta t \to 0}\sigma^2 \Delta t/2$. 

%%%%%%%%%%%%%%%%%%%%%%%%%%%%%%%%%%%%%%%%%%%%%%%%%%%%%%%%%%%%%%%%%%%%%%%%%%%%%%%%
%%%%%%%%%%%%%%%%%%%%%%%%%%%%%%%%%%%%%%%%%%%%%%%%%%%%%%%%%%%%%%%%%%%%%%%%%%%%%%%%
%%%%%%%%%%%%%%%%%%%%%%%%%%%%%%%%%%%%%%%%%%%%%%%%%%%%%%%%%%%%%%%%%%%%%%%%%%%%%%%%
\subsubsection{The propagator prior to first reset.}
\l{sec:propagator}
What is the probability of occurrence of particle trajectories that start at position
$x_0$ and end at a given location $x$ at time $t=N\Delta t$ without having undergone any
reset event? From the discrete-time dynamics given by Eq.
(\ref{eq:dynamics}) and the joint distribution
(\ref{eq:joint-distribution}), the probability of
occurrence of a given particle trajectory $\{x_i\}_{0 \le i \le N}\equiv\{x_0,x_1,x_2,\ldots,x_{N-1},x_N=x\}$
is given by
\bea
&&P_{\rm{no\;res}}[\{x_i\}]={\rm det}({\cal J})\left(\frac{1}{2\pi
\sigma^2}\right)^{N/2}\nonumber\\
&&\times \prod_{i=1}^{N}\exp\!\left(-\frac{(x_i-x_{i-1}-\mu
\bar{F}(x_i)\Delta t)^2}{2\sigma^2(\Delta t)^2}\right)\!\prod_{i=0}^{N-1}\Big(1-r(x_i)\Delta t\Big).\nonumber\\
\eea
Here, the factor $\prod_{i=0}^{N-1}\left(1-r(x_i)\Delta t\right)$ enforces the condition that the particle
has not reset at any of the instants $t_i,~i=0,1,2,\ldots,N-1$, while ${\cal J}$ is the Jacobian matrix  for the transformation $\{\eta_i\}\rightarrow\{x_i\}$, which
is obtained from Eq. (\ref{eq:dynamics}) as ${\cal J}_{1\leq i,j\leq N}\equiv\left(\frac{\partial\eta_i}{\partial x_j}\right)  $ or equivalently
\be
{\cal J}=\left(\begin{array}{cccc}
\frac{1}{\Delta t}-\frac{\mu F'(x_1)}{2} & 0 & 0 & \ldots\\
-\frac{1}{\Delta t}-\frac{\mu F'(x_1)}{2} & \frac{1}{\Delta
t}-\frac{\mu F'(x_2)}{2} & 0 & \ldots\\
\vdots & \vdots & \vdots & \vdots
\end{array}\right)_{N\times N},
\ee
with primes denoting derivative with respect to $x$. One thus has 
\bea
&&{\rm det}({\cal J})=\left(\frac{1}{\Delta t}\right)^N \,\prod_{i=1}^{N}\left(1-\frac{\Delta t \mu F'(x_i)}{2}\right)\nonumber\\
&&\simeq\left(\frac{1}{\Delta t}\right)^{N}\exp\left(-\sum_{i=1}^{N}\frac{\Delta t \mu F'(x_i)}{2}\right),
\eea
where in obtaining the last step, we have used the smallness of
$\Delta t$. Thus, for small $\Delta t$,  we get 
\bea
&&P_{\rm{no\;res}}[\{x_i\}]=
\left(\frac{1}{2\pi \sigma^2(\Delta
t)^2}\right)^{N/2}\nonumber\\
&&\times\prod_{i=1}^{N}\exp\left(-\frac{(x_i-x_{i-1}-\mu\bar{F}(x_i)\Delta
t)^2}{2\sigma^2(\Delta t)^2}-\frac{\Delta t \mu
F'(x_i)}{2}\right)\nonumber \\
&&\times\prod_{i=0}^{N-1}\exp\Big(-r(x_i)\Delta t\Big)\nonumber\\
&&=\left(\frac{1}{2\pi \sigma^2(\Delta
t)^2}\right)^{N/2}\exp\Big(\Delta
 t\Big[r(x_N)-r(x_0)\Big]\Big)\nonumber\\
&&\times \exp\Big(-\Delta
t\sum_{i=1}^{N}\Big[\frac{[(x_i-x_{i-1}-\mu\bar{F}(x_i)\Delta t)/\Delta
t]^2}{2\sigma^2\Delta t}\nonumber\\
&& +\frac{\mu F'(x_i)}{2}+r(x_i)\Big]\Big). \nonumber \\
\l{eq:path-0} 
\eea

From Eq. (\ref{eq:path-0}), it follows by considering all possible
trajectories that the
probability density that the particle while starting at position $x_0$ ends at
a given location $x$ at time $t=N\Delta t$ without having undergone any reset
event is given by
\bea
&&\hspace{-0.6cm}P_{\rm no\;res}(x,t|x_0,0)=\left(\frac{1}{2\pi \sigma^2(\Delta
t)^2}\right)^{N/2}\!\exp\Big(\Delta
t\Big[r(x)-r(x_0)\Big]\Big)\nonumber \\
&&\hspace{-0.6cm}\times\prod_{i=1}^{N-1}\int_{-\infty}^{\infty}{\rm
d}x_i\exp\Big(-\Delta
t\sum_{i=1}^{N}\Big[\frac{[(x_i-x_{i-1}-\mu \bar{F}(x_i)\Delta
t)/\Delta t]^{2}}{2\sigma^2 \Delta t}\nonumber\\
\hspace{-0.6cm}&&+\frac{\mu F'(x_i)}{2}+r(x_i)\Big]\Big).
\eea
In the limit of continuous time, defining
${\cal D}x(t)\equiv\lim_{N\to\infty}\Big(\frac{1}{4\pi D\Delta t}\Big)^{N/2}\prod_{i=1}^{N-1}\int_{-\infty}^{\infty}{\rm
d}x_i,$
 one gets the exact expression for the corresponding probability density
 as the following path integral:
\bea
P_{\rm no\;res}(x,t|x_0,0)= \int_{x(0)=x_0}^{x(t)=x}{\cal
D}x(t)\exp\left(-S_{\rm res}[\{x(t)\}]\right),
\l{eq:P-no-reset-0}\nonumber\\
\eea
where on the right hand side of Eq.~(\ref{eq:P-no-reset-0}), we have introduced the {\em resetting action} as
\be
\hspace{-0.2cm}S_{\rm res}[\{x(t)\}]=\int_0^{t}{\rm
d}t\left[\frac{[({\rm d}x/{\rm d}t)-\mu F(x) )^{2}}{4D}+\frac{\mu
F'(x)}{2}+r(x)\right].
\ee
Invoking the Feynman-Kac
formalism, we identify the path integral on the
right hand side of Eq.~(\ref{eq:P-no-reset-0}) with the propagator of a
quantum mechanical evolution in (negative) imaginary time due to a quantum
Hamiltonian $H_{\rm q}$ (see Appendix), to get
\be
P_{\rm no\;res}(x,t|x_0,0)=\exp\left(\frac{\mu}{2D} \int_{x_0}^x F(x)~{\rm
d}x\right)G_{\rm q}(x,-it|x_0,0), 
\l{eq:pnoresq}
\ee
with
\be
G_{\rm q}(x,-it|x_0,0)\equiv\langle
x|\exp(-H_{\rm q}t)|x_0\rangle,
\l{eq:qma}
\ee
where the quantum Hamiltonian is 
\be
H_{\rm q}\equiv-\frac{1}{2m_{\rm q}}\frac{\partial^{2}}{\partial x^{2}}+V_{\rm q}(x),
\l{eq:quantum-hamiltonian}
\ee
the mass in the equivalent quantum problem is 
\be
m_{\rm q}\equiv\frac{1}{2D},
\ee
and the quantum potential is given by
\be
V_{\rm q}(x)\equiv \frac{\mu^2(F(x))^2}{4D}+\frac{\mu F'(x)}{2}+r(x).
\l{eq:quantum-potential}
\ee
Note that in the quantum propagator in Eq.~(\ref{eq:qma}), the Planck's
constant has been set to unity, $\hbar=1$, while the time $\tau$ of
propagation is imaginary: $\tau=-it$~\cite{Wick}. 
Since the Hamiltonian contains no explicit time dependence,
the propagator $G_{\rm q}(x,-it|x_0,0)$ is effectively a function of the time $t$ to
propagate from the initial location $x_0$ to the final location $x$, and
not individually of the initial and final times.  
Let us note that on using
$D=k_{\rm B}T \mu$, the prefactor equals
$\exp\left(-Q(t)/2k_{\rm B}T\right)$, where $Q(t) \equiv \int_{x_0}^{x}
\partial_x V(x)\,{\rm d}x$ is the heat absorbed by the particle from the environment along the trajectory $\{x(t)\}$~\cite{Sekimoto:1998,Sekimoto:2000}.

%%%%%%%%%%%%%%%%%%%%%%%%%%%%%%%%%%%%%%%%%%%%%%%%%%%%%%%%%%%%%%%%%%%%%%%%%%%%%%%%
%%%%%%%%%%%%%%%%%%%%%%%%%%%%%%%%%%%%%%%%%%%%%%%%%%%%%%%%%%%%%%%%%%%%%%%%%%%%%%%%
%%%%%%%%%%%%%%%%%%%%%%%%%%%%%%%%%%%%%%%%%%%%%%%%%%%%%%%%%%%%%%%%%%%%%%%%%%%%%%%%
\subsubsection{Distribution of the first-reset time}
\l{sec:prob-first-reset}

Let us now ask for the probability of occurrence of trajectories that start at position
$x_0$ and reset for the first time at time $t$. In terms of $P_{\rm
no\;res}(x,t|x_0,0)$, one gets this probability density as  
\be
P_{\rm res}(t|x_0)=\int_{-\infty}^{\infty}{\rm d}y~r(y)P_{\rm
no\;res}(y,t|x_0,0),
\l{eq:Pt-final}
\ee
since by the very definition of $P_{\rm res}(t|x_0)$, a reset has to happen only at
the final time $t$ when the particle has reached the location $y$, where
$y$ may in principle take any value in the
interval $[-\infty,\infty]$. The probability density $P_{\rm res}(t|x_0)$ is
normalized as $\int_0^\infty {\rm d}t~P_{\rm res}(t|x_0)=1$.

%%%%%%%%%%%%%%%%%%%%%%%%%%%%%%%%%%%%%%%%%%%%%%%%%%%%%%%%%%%%%%%%%%%%%%%%%%%%%%%%
%%%%%%%%%%%%%%%%%%%%%%%%%%%%%%%%%%%%%%%%%%%%%%%%%%%%%%%%%%%%%%%%%%%%%%%%%%%%%%%%
%%%%%%%%%%%%%%%%%%%%%%%%%%%%%%%%%%%%%%%%%%%%%%%%%%%%%%%%%%%%%%%%%%%%%%%%%%%%%%%%
\subsubsection{Spatial time-dependent probability distribution}
\l{sec:prob-distr}
Using renewal theory, we now show that knowing $P_{\rm no\;res}(x,t|x_0,0)$ and $P_{\rm res}(t|x_0)$ is sufficient to
obtain the spatial distribution of the particle at any time $t$.  
The probability density that the particle is at $x$ at time $t$ while
starting from $x_0$ is given by
\bea
&&P(x,t|x_0,0)=P_{\rm no\;res}(x,t|x_0,0)\nonumber\\
&&+\int_0^t {\rm
d}\tau\int_{-\infty}^\infty {\rm
d}y~r(y)P(y,t-\tau|x_0,0)P_{\rm no\;res}(x,t|x_0,t-\tau)\nonumber \\
&&=P_{\rm no\;res}(x,t|x_0,0)\nonumber\\
&& +\int_0^t {\rm
d}\tau~R(t-\tau|x_0)P_{\rm no\;res}(x,t|x_0,t-\tau), 
\l{eq:Pxt-final}
\eea
where we have defined the probability density to reset at time $t$ as
\be
R(t|x_0)\equiv \int_{-\infty}^\infty {\rm
d}y~r(y)P(y,t|x_0,0).
\l{eq:ft-definition}
\ee
One may easily understand Eq.~(\ref{eq:Pxt-final}) by invoking the
theory of renewals \cite{Cox:1962} and realizing that the dynamics is
renewed each time the particle resets to $x_0$. This may be seen as
follows. The
particle while starting from $x_0$ may reach $x$ at time $t$ by
experiencing not a single reset; the corresponding contribution to the
spatial distribution is given by the first term on the right hand side of Eq.~(\ref{eq:Pxt-final}). The particle may also reach $x$ at time $t$ by experiencing
the last reset event (i.e., the last renewal) at time instant $t-\tau$, with $\tau \in
[0,t]$, and then propagating from the reset location $x^{(\rm r)}=x_0$ to
$x$ without experiencing any further reset, where the last reset may take place with rate $r(y)$ from any
location $y \in [-\infty,\infty]$ where the particle happened to be at
time $t-\tau$; such contributions are represented by the second term on
the right hand side of Eq.~(\ref{eq:Pxt-final}). The spatial distribution is
normalized as $\int_{-\infty}^\infty {\rm
d}x~P(x,t|x_0,0)=1$ for all possible values of $x_0$ and $t$.

Multiplying both sides of Eq. (\ref{eq:Pxt-final}) by $r(x)$, and then integrating over $x$, we get
\bea
&&R(t|x_0)=\int_{-\infty}^\infty {\rm d}x~r(x)P_{\rm no\;res}(x,t|x_0,0)\nonumber\\
&&+\int_0^t {\rm
d}\tau\left[\int_{-\infty}^\infty {\rm
d}x~r(x)P_{\rm no\;res}(x,t|x_0,t-\tau)\right]\! R(t-\tau|x_0). \nonumber\\
\eea
The square-bracketed quantity on the right hand side is nothing but $P_{\rm res}(\tau|x_0)$,
so that we get
\be
R(t|x_0)=P_{\rm res}(t|x_0)+\int_0^t {\rm d}\tau~P_{\rm res}(\tau|x_0)R(t-\tau|x_0).
\l{eq:Rt-definition}
\ee
Taking the Laplace transform on both sides of Eq.~(\ref{eq:Rt-definition}), we get
\be
\widetilde{R}(s|x_0)=\widetilde{P}_{\rm res}(s|x_0)+\widetilde{P}_{\rm res}(s|x_0)\widetilde{R}(s|x_0),
\l{eq:Rs-equation}
\ee
where $\widetilde{R}(s|x_0)$ and $\widetilde{P}_{\rm res}(s|x_0)$ are respectively the Laplace transforms of $R(t|x_0)$ and
$P_{\rm res}(t|x_0)$. Solving for $\widetilde{R}(s|x_0)$ from 
Eq.~(\ref{eq:Rs-equation}) yields
\be
\widetilde{R}(s|x_0)=\frac{\widetilde{P}_{\rm res}(s|x_0)}{1-\widetilde{P}_{\rm res}(s|x_0)}.
\l{eq:Rs}
\ee
Next, taking the Laplace transform with respect to time on both sides of Eq. (\ref{eq:Pxt-final}), we obtain
\bea
\widetilde{P}(x,s|x_0,0)&=&\Big(1+\widetilde{R}(s|x_0)\Big)\widetilde{P}_{\rm
no\;res}(x,s|x_0) \nonumber \\
&=&\frac{\widetilde{P}_{\rm
no\;res}(x,s|x_0)}{1-\widetilde{P}_{\rm res}(s|x_0)}, 
\l{eq:Pxs}
\eea
where we have used Eq. (\ref{eq:Rs}) to obtain the last equality. An
inverse Laplace transform of Eq.~(\ref{eq:Pxs}) yields the time-dependent spatial
distribution $P(x,t|x_0,0)$. 

\subsubsection{Stationary spatial distribution}
\l{sec:stat-distr}
On applying the final value
theorem, one may obtain the stationary spatial distribution as
\be
P_{\rm st}(x|x_0)=\lim_{s\to 0}s\widetilde{P}(x,s|x_0,0)=\lim_{s\to 0}s\frac{\widetilde{P}_{\rm
no\;res}(x,s|x_0)}{1-\widetilde{P}_{\rm res}(s|x_0)},
\l{eq:Pst-0}
\ee
provided the stationary distribution (i.e., $\lim_{t \to \infty}P(x,t|x_0,0)$) exists.
Now, since $P_{\rm res}(t|x_0)$ is normalized to unity, $\int_0^\infty {\rm
d}t~P_{\rm res}(t|x_0)=1$, we may expand its Laplace transform to
leading orders in $s$ as
$\widetilde{P}_{\rm res}(s|x_0)\equiv \int_0^\infty {\rm
d}t~\exp(-st)P_{\rm res}(t|x_0)=1-s\langle t \rangle_{\rm res}+O(s^2)$, provided that the mean first-reset time $\langle t
\rangle_{\rm res}$, defined as
\be
\langle t
\rangle_{\rm res} \equiv \int_0^\infty {\rm d}t~tP_{\rm res}(t|x_0),
\ee
is finite. Similarly, we may expand $\widetilde{P}_{\rm
no\;res}(x,s|x_0,0)$ to leading orders in $s$ as $\widetilde{P}_{\rm
no\;res}(x,s|x_0,0)=\int_0^\infty {\rm d}t~P_{\rm no\;res}(x,t|x_0,0)-s\int_0^\infty
{\rm d}t~tP_{\rm no\;res}(x,t|x_0,0)+O(s^2)$, provided that $\int_0^\infty
{\rm d}t~tP_{\rm no\;res}(x,t|x_0,0)$ is finite. From Eq.~(\ref{eq:Pst-0}), we thus find the stationary spatial distribution
to be given by the integral over all times of the propagator prior to
first reset divided by the mean first-reset time:
\be
P_{\rm st}(x|x_0)=\frac{1}{\langle t \rangle_{\rm res}}\displaystyle\int_0^\infty {\rm d}t~P_{\rm
no\;res}(x,t|x_0,0).
\l{eq:Pxstat-final}
\ee

%%%%%%%%%%%%%%%%%%%%%%%%%%%%%%%%%%%%%%%%%%%%%%%%%%%%%%%%%%%%%%%%%%%%%%%%%%%%%%%%
%%%%%%%%%%%%%%%%%%%%%%%%%%%%%%%%%%%%%%%%%%%%%%%%%%%%%%%%%%%%%%%%%%%%%%%%%%%%%%%%
\section{EXACTLY SOLVED EXAMPLES}
\l{sec:applications}

%%%%%%%%%%%%%%%%%%%%%%%%%%%%%%%%%%%%%%%%%%%%%%%%%%%%%%%%%%%%%%%%%%%%%%%%%%%%%%%%
%%%%%%%%%%%%%%%%%%%%%%%%%%%%%%%%%%%%%%%%%%%%%%%%%%%%%%%%%%%%%%%%%%%%%%%%%%%%%%%%
%%%%%%%%%%%%%%%%%%%%%%%%%%%%%%%%%%%%%%%%%%%%%%%%%%%%%%%%%%%%%%%%%%%%%%%%%%%%%%%%
\subsection{Free particle with space-independent resetting}
\l{subsec:constant-resetting}

Let us first consider the simplest case of free diffusion with a space-independent
rate of resetting $r(x)=r$, with $r$ a positive constant having the
dimension of inverse time. Here, on using Eq.~(\ref{eq:pnoresq}) with $F(x)=0$, we have
\bea
P_{\rm no\;res}(x,t|x_0,0)&=&G_{\rm q}(x,-it|x_0,0)\nonumber\\
&=&\langle
x|\exp(-H_{\rm q}t)|x_0\rangle,
\l{eq:Pnoreset-free-parabola}
\eea
where the quantum Hamiltonian is in this case, following
Eqs.~(\ref{eq:quantum-hamiltonian}-\ref{eq:quantum-potential}), given by
\be
H_{\rm q}=-\frac{1}{2m_{\rm q}}\frac{\partial^{2}}{\partial x^{2}} + r;~~m_{\rm q}=\frac{1}{2D},~\hbar=1.
\ee
Since in the present situation, the effective quantum potential $V_{\rm q}(x)=r$ is
space independent, we may rewrite Eq.~(\ref{eq:Pnoreset-free-parabola}) as:
\bea
P_{\rm no\;res}(x,t|x_0,0)&=&\exp(-rt)G_{\rm q}(x,-it|x_0,0),
\l{eq:Pnoreset-constant-resetting-0}
\eea
with
\be
G_{\rm q}(x,-it|x_0,0)\equiv\langle
x|\exp(-H_{\rm q}t)|x_0\rangle,
\ee
where the quantum Hamiltonian is now that of a free particle:
\be
H_{\rm q}\equiv-\frac{1}{2m_{\rm
q}}\frac{\partial^{2}}{\partial x^{2}};~~m_{\rm q}=\frac{1}{2D},~\hbar=1.  
\l{eq:Hq-free}
\ee
Therefore, the statistics of resetting of a free particle under a
space-independent rate of resetting may be found from the quantum propagator of a free particle, which is given by~\cite{Schulman:1981}
\be
G_{\rm q}(x,\tau|x_0,0)=\sqrt{\frac{m_{\rm q}}{2\pi \hbar i  \tau }}
\exp \left(-\frac{m_{\rm q}(x-x_0)^2}{2\hbar i\tau}\right). 
\l{eq:G-free-parabola}
\ee
Plugging in Eq.~(\ref{eq:G-free-parabola}) the parameters in
Eq.~(\ref{eq:Hq-free}) together with $\tau=-it$, we have
\be
G_{\rm q}(x,-it|x_0,0)=\frac{1}{\sqrt{4\pi Dt}}\exp\left(-\frac{(x-x_0)^2}{4Dt}\right).
\l{eq:G-free-parabola-final}
\ee
Using Eq.~(\ref{eq:G-free-parabola-final}) in Eq.~(\ref{eq:Pnoreset-constant-resetting-0}), we thus obtain
\be
P_{\rm no\;res}(x,t|x_0,0)=\frac{\exp(-rt)}{\sqrt{4\pi
Dt}}\exp\left(-\frac{(x-x_0)^2}{4Dt}\right),
\l{eq:Pnoreset-constant-resetting}
\ee
and hence, the distribution of the first-reset time may be found on
using Eq. (\ref{eq:Pt-final}):
\bea
P_{\rm res}(t|x_0)&=& r \exp(-rt)\frac{1}{\sqrt{4\pi Dt}}\int_{-\infty}^{\infty}{\rm d}x~
\!\exp\left(-\frac{(x-x_0)^2}{4Dt}\right)\nonumber\\
&=& r\exp(-rt),
\l{eq:Pt-constant-resetting}
\eea
which is normalized to unity: $\int_0^{\infty}{\rm d}t~P_{\rm res}(t|x_0)=1$, as expected. 

Using Eq. (\ref{eq:Pt-constant-resetting}), we get $\widetilde{P}_{\rm res}(s|x_0)=r/(s+r)$, so 
that Eq. (\ref{eq:Rs}) yields $\widetilde{R}(s|x_0)=r/s$. An inverse Laplace transform yields  $R(t|x_0)=r$, as
also follows from Eq.~(\ref{eq:ft-definition}) by substituting $r(y)=r$
and noting that
$P(y,t|x_0,0)$ is normalized with respect to $y$.

Next, the probability density that the particle is at $x$ at time $t$, while starting
from $x_0$, is obtained on using Eq. (\ref{eq:Pxt-final}) as
\bea
P(x,t|x_0,0)&=&\frac{\exp(-rt)}{\sqrt{4\pi
D\tau}}\exp(-(x-x_0)^2/(4Dt))\nonumber\\
&+&r \int_0^t {\rm
d}\tau~\frac{\exp(-r\tau)}{\sqrt{4\pi
D\tau}}\exp(-(x-x_0)^2/(4D\tau)).\nonumber\\
\l{eq:Pxt-constant-resetting}
\eea
Taking the limit $t \to \infty$, we obtain the
stationary spatial distribution as
\bea
P_{\rm st}(x|x_0)&=& r \int_0^\infty {\rm
d}\tau~\exp(-r\tau)
\frac{\exp(-(x-y)^2/(4D\tau))}{\sqrt{4\pi
D\tau}},\nonumber\\
\l{eq;Pxstat-constant-resetting}
\eea
which may also be obtained by using Eqs. (\ref{eq:Pxstat-final}) and
(\ref{eq:Pnoreset-constant-resetting}), and also Eq.
(\ref{eq:Pt-constant-resetting}) that implies that $\langle t
\rangle_{\rm res}=1/r$.
From Eq.~(\ref{eq:Pxt-constant-resetting}), we obtain an exact expression for
the time-dependent spatial distribution as
\bea
&&P(x,t|x_0,0)=\frac{\exp(-rt)\exp(-(x-x_0)^2/4Dt)}{\sqrt{4\pi Dt}}\nonumber\\
&&+\frac{\exp\left(-\frac{|x-x_0|}{\sqrt{D/r}}\right){\rm
erfc}\left(\frac{|x-x_0|}{\sqrt{4Dt}}-\sqrt{rt}\right)}{\sqrt{4D/r}} \nonumber\\
&&-\frac{\exp\left(\frac{|x-x_0|}{\sqrt{D/r}}\right){\rm
erfc}\left(\frac{|x-x_0|}{\sqrt{4D
t}}+\sqrt{rt}\right)}{\sqrt{4D
t}},
\eea
while Eq.~(\ref{eq;Pxstat-constant-resetting}) yields the exact
stationary distribution as
\bea
P_{\rm
st}(x|x_0)=\frac{1}{2\sqrt{D/r}}\exp\left(-\frac{|x-x_0|}{\sqrt{D/r}}\right),
\l{eq:Pxstat-constant-resetting-final}
\eea
where ${\rm erfc}(x)\equiv (2/\sqrt{\pi})\int_x^\infty {\rm d}t~\exp(-t^2)$ is the complementary error function.
The stationary distribution (\ref{eq:Pxstat-constant-resetting-final})
may be put in the scaling form
\be
P_{\rm st}(x|x_0)=\frac{1}{2\sqrt{D/r}}{\cal R}\left(\frac{|x-x_0|}{\sqrt{D/r}}\right),
\l{eq:Pxstat-final-rconstant}
\ee
where the scaling function is given by ${\cal R}(y)\equiv\exp(-y)$.
For the particular case $x_0=0$,
Eq.~(\ref{eq:Pxstat-constant-resetting-final}) matches with the result derived in Ref.~\cite{Evans:2011-1}.
Note that the steady state distribution (\ref{eq:Pxstat-final-rconstant}) exhibits a cusp at the
resetting location $x_0$. Since the resetting location is taken to be
the same as the initial location, the particle visits repeatedly in time the initial location, thereby keeping a memory of
the latter that makes an explicit appearance even in the long-time
stationary state.

%%%%%%%%%%%%%%%%%%%%%%%%%%%%%%%%%%%%%%%%%%%%%%%%%%%%%%%%%%%%%%%%%%%%%%%%%%%%%%%%
%%%%%%%%%%%%%%%%%%%%%%%%%%%%%%%%%%%%%%%%%%%%%%%%%%%%%%%%%%%%%%%%%%%%%%%%%%%%%%%%
%%%%%%%%%%%%%%%%%%%%%%%%%%%%%%%%%%%%%%%%%%%%%%%%%%%%%%%%%%%%%%%%%%%%%%%%%%%%%%%%
\subsection{Free particle with  ``parabolic'' resetting}
\l{subsec:parabolic-resetting}

We now study the dynamics of a free Brownian particle whose position is reset to the initial position $x_0$ with a rate of resetting that is proportional to the square of the current position of the particle. 
In this case, we have $r(x)=\alpha x^2$, with $\alpha>0$ having the dimension of $1/(({\rm Length})^{2}{\rm Time})$. 
From Eqs. (\ref{eq:pnoresq}) and (\ref{eq:qma}), and given that
in this case $F(x)=0$, we get 
\bea
P_{\rm no\;res}(x,t|x_0,0)=G_{\rm q}(x,-it|x_0,0) = \langle
x|\exp(-H_{\rm q}t)|x_0\rangle,\nonumber\\
\l{eq:parabolic-resetting-Pxt-0-x0}
\eea
with the Hamiltonian obtained from Eq. (\ref{eq:quantum-hamiltonian}) by
setting $V_{\rm q}(x)=\alpha x^2$:
\be
H_{\rm q}= -\frac{1}{2m_{\rm q}}\frac{\partial^{2}}{\partial
x^{2}} + \alpha x^2;~~m_{\rm q}=\frac{1}{2D},~\hbar=1. 
\l{eq:Hq-free-parabola}
\ee
We thus see that the statistics of resetting of a free particle subject
to a ``parabolic" rate of resetting may be found from the propagator of a quantum harmonic oscillator.
Following Schulman~\cite{Schulman:1981}, a quantum harmonic oscillator with the Hamiltonian given by
\be
H_{\rm q}=-\frac{1}{2m_{\rm q}}\frac{\partial^{2}}{\partial x^{2}} +
\frac{1}{2}m_{\rm q} \omega_{\rm q}^2 x^2,
\ee
with $m_{\rm q}$ and $\omega_{\rm q}$ being the mass and the frequency
of the oscillator, has the quantum propagator
\bea
&&\hspace{-0.5cm}G_{\rm q}(x,\tau|x_0,0)=\nonumber\\
&&\hspace{-0.5cm}\sqrt{\frac{m_{\rm q}\omega_{\rm q}}{2i\pi\hbar  \sin \omega_{\rm q} \tau }} \exp\left(\frac{i\omega_{\rm q}}{2\hbar\sin \omega_{\rm q} \tau}  [   (x^2 + x_0^2)\cos \omega_{\rm q} \tau - 2 x x_0]    \right).\nonumber\\
\l{eq:Gq-free-parabola}
\eea
Using the parameters given in Eq.~(\ref{eq:Hq-free-parabola}), and
substituting $\tau=-it$ and $\omega_{\rm q}=\sqrt{4D\alpha}$ in
Eq.~(\ref{eq:Gq-free-parabola}), we have
\bea
&&\hspace{-0.4cm}G_{\rm q}(x,-it|x_0,0)=\frac{(\alpha/D)^{1/4}}{\sqrt{2\pi\sinh(t\sqrt{4D\alpha})}}\nonumber\\
&&\hspace{-0.4cm}\times\exp\left(-\frac{\sqrt{\alpha/D}}{2\sinh(t\sqrt{4D\alpha})}[(x_0^{2}+y^{2})\cosh(t\sqrt{4D\alpha})-2x x_0]\right).\nonumber\\
\l{eq:Gq-free-parabola-1}
\eea

We may now derive the statistics of resetting by using the
propagator~(\ref{eq:Gq-free-parabola-1}).
Equation~(\ref{eq:parabolic-resetting-Pxt-0-x0}) together with
Eq.~(\ref{eq:Gq-free-parabola-1}) imply
\bea
&&P_{\rm
no\;res}(x,t|x_0,0)=\frac{(\alpha/D)^{1/4}}{\sqrt{2\pi\sinh(t\sqrt{4D\alpha})}}\nonumber\\
&&\hspace{-0.4cm}\times\exp\left(-\frac{\sqrt{\alpha/D}}{2\sinh(t\sqrt{4D\alpha})}[(x_0^{2}+x^{2})\cosh(t\sqrt{4D\alpha})-2x_0x]\right).\nonumber\\
\l{eq:parabolic-resetting-Pxt-x0}
\eea
Integrating Eq.~(\ref{eq:parabolic-resetting-Pxt-x0}) over $x$, we get
the distribution of the first-reset time as
\bea
&&P_{\rm res}(t|x_0) =\int_{-\infty}^{\infty}{\rm d}y~r(y)P_{\rm
no\;res}(y,t|x_0,0)\nonumber \\
&&=\frac{\alpha(\alpha/D)^{1/4}}{\sqrt{\sinh(t\sqrt{4D\alpha})}}\exp\left(-x_0^2\frac{\sqrt{\alpha/D}}{2}\tanh(t\sqrt{4D\alpha})\right)\nonumber\\
&&\times \frac{\sqrt{\alpha/D}\coth(t\sqrt{4D\alpha})+x_0^2(\alpha/D){\rm
cosech}^2(t\sqrt{4D\alpha})}{(\alpha/D)^{5/4}\coth^{5/2}(t\sqrt{4D\alpha})}.\nonumber\\
\l{eq:parabolic-resetting-Pt-x0}
\eea

For the case $x_0=0$, Eqs. (\ref{eq:parabolic-resetting-Pxt-x0}) and
(\ref{eq:parabolic-resetting-Pt-x0}) reduce to simpler expressions:
\bea
&&P_{\rm
no\;res}(x,t|x_0=0,0)=\nonumber\\
&&\frac{(\alpha/D)^{1/4}}{\sqrt{2\pi\sinh(t\sqrt{4D\alpha})}}\exp\left(-\frac{x^2\sqrt{\alpha/D}\coth(t\sqrt{4D\alpha})}{2}\right),\nonumber\\
\l{eq:parabolic-resetting-Pxt-x00}
\eea
and
\bea
P_{\rm res}(t|x_0=0)
=\sqrt{D\alpha}\frac{(\tanh(t\sqrt{4D\alpha}))^{3/2}}{\sqrt{\sinh(t\sqrt{4D\alpha})}}.
\l{eq:parabolic-resetting-Pt-x00}
\eea
Equation (\ref{eq:parabolic-resetting-Pt-x00}) may be put in the scaling form
\be
P_{\rm res}(t|x_0=0)=\sqrt{D\alpha}\,{\cal G}\!\left(t\sqrt{4D\alpha}\right),
\ee
with ${\cal G}(y) = \tanh(y)^{3/2}/\sqrt{\sinh(y)}$. Equation~(\ref{eq:parabolic-resetting-Pt-x00}) yields the mean first-reset time $\langle t \rangle_{\rm res}$ for $x_0=0$  to be given by
\be
\langle t \rangle_{\rm res}=\frac{(\Gamma(1/4))^2}{4\sqrt{2\pi
D\alpha}},
\l{eq:tav-x00}
\ee
where $\Gamma$ is the Gamma function.  Equations (\ref{eq:parabolic-resetting-Pt-x00}) and
(\ref{eq:tav-x00}) yield the stationary spatial
distribution on using Eq.~(\ref{eq:Pxstat-final}):
\bea
&&P_{\rm st}(x|x_0=0)=\frac{4\sqrt{
D\alpha}(\alpha/D)^{1/4}}{(\Gamma(1/4))^2}\nonumber\\
&&\times \displaystyle\int_0^\infty \frac{{\rm
d}t}{\sqrt{\sinh(t\sqrt{4D\alpha})}}\exp\left(-\frac{x^2\sqrt{\alpha/D}\coth(t\sqrt{4D\alpha})}{2}\right).\nonumber\\
&&=\frac{2^{3/4}(\alpha/D)^{1/4}}{\sqrt{\pi}\Gamma(1/4)}\left(\frac{x^{2}\sqrt{\alpha/D}}{2}\right)^{1/4}K_{1/4}\left(\frac{x^{2}\sqrt{\alpha/D}}{2}\right),
\l{eq:stationary-px-parabolic-resetting}
\eea
where $K_n(x)$ is the $n-$th order modified Bessel function of the second kind.
Equation~(\ref{eq:stationary-px-parabolic-resetting}) implies that the
stationary distribution is symmetric around $x=0$, which is expected
since the resetting rate is symmetric around $x_0=0$.
The stationary distribution (\ref{eq:stationary-px-parabolic-resetting})
may be put in the scaling form
\be
P_{\rm
st}(x|x_0=0)=\frac{2^{3/4}(\alpha/D)^{1/4}}{\sqrt{\pi}\Gamma(1/4)}{\cal
R}\left(\frac{x}{(D/\alpha)^{1/4}}\right),
\ee
where the scaling function is given by ${\cal R}(y)=(y^2/2)^{1/4}K_{1/4}(y^2/2)$.

\begin{figure}[h]
\centering
\includegraphics[width=0.4\textwidth]{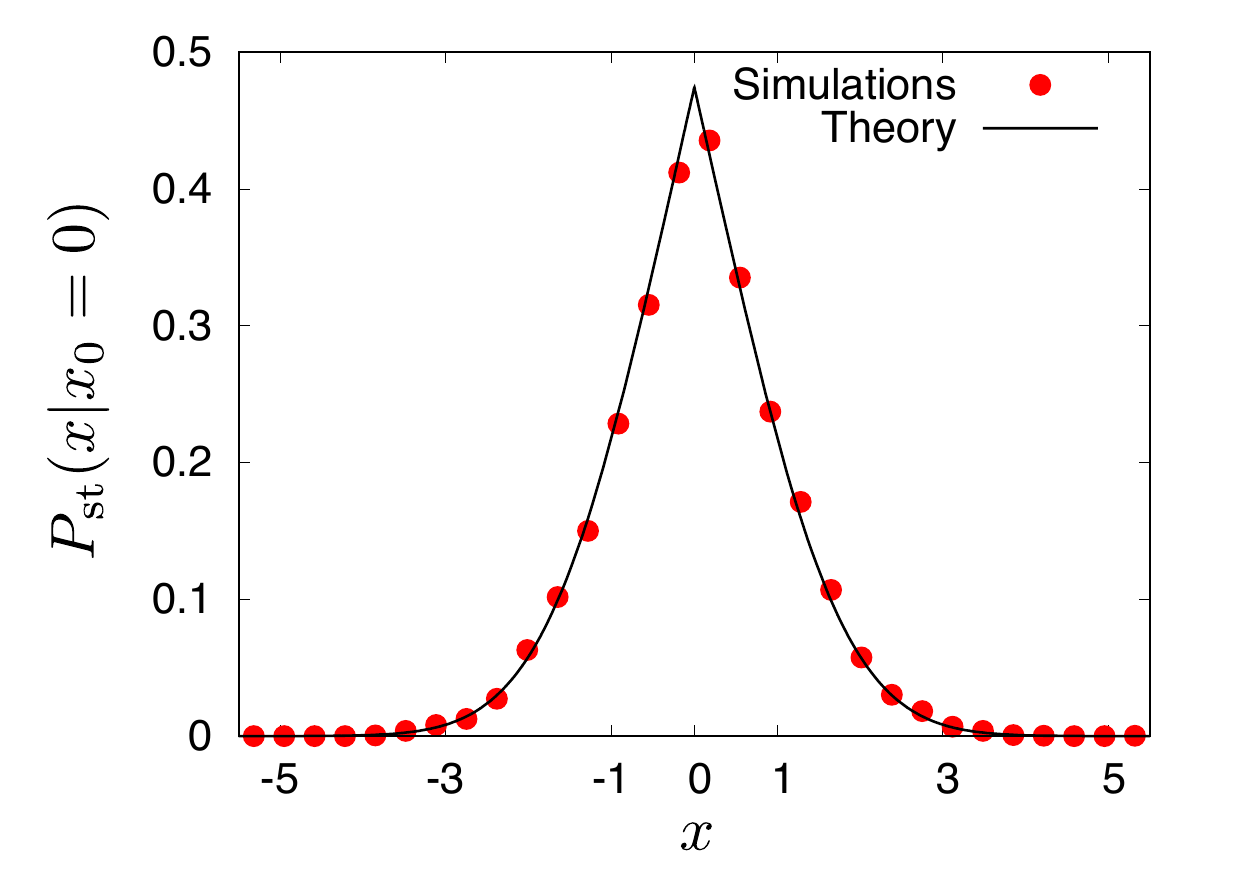}
\caption{\textbf{Theory versus simulation results for the stationary
spatial distribution of a free Brownian particle undergoing parabolic
resetting.} The points denote simulation results, while the line stands
for the exact analytical expression given in Eq.
(\ref{eq:stationary-px-parabolic-resetting}). The numerical results were obtained from 
$10^4$ independent simulations of the Langevin dynamics described in Section
\ref{sec:model}. The error bars associated with the data points are smaller than the symbol
size. The parameter values are
$D=1.0,\alpha=0.5$.}
\label{fig:free-parabola}
\end{figure}

The result (\ref{eq:stationary-px-parabolic-resetting}) is checked in
simulations in Fig. \ref{fig:free-parabola}. The simulations involved
numerically integrating the dynamics described in Section
\ref{sec:model}, with integration timestep equal to $0.01$. 
Using $\int_0^\infty {\rm
d}t~t^{\mu-1}K_\nu(t)=2^{\mu-2}\Gamma\Big(\frac{\mu}{2}-\frac{\nu}{2}\Big)\Gamma\Big(\frac{\mu}{2}+\frac{\nu}{2}\Big)$
for $|{\rm Re}~\nu|<|{\rm Re}~\mu|$ \cite{Olver:2016}, we find
that $P_{\rm st}(x|x_0=0)$ given by Eq.~(\ref{eq:stationary-px-parabolic-resetting}) is correctly normalized to
unity. Moreover, using the results that as $x\to 0$, we have
$K_\nu(x)=\frac{\Gamma(\nu)}{2}\Big(\frac{x}{2}\Big)^{-\nu}$ for ${\rm
Re}~\nu>0$ and that as $x\to \infty$, we have
$K_\nu(x)=\Big(\frac{\pi}{2x}\Big)^{1/2}\exp(-x)$ for real $x$
\cite{Olver:2016}, we
get
\bea
P_{\rm st}(x|x_0=0) \sim
\left\{
\begin{array}{ll}
\frac{(\alpha/D)^{1/4}}{\sqrt{\pi}} & ~~{\rm for}~~x\to 0,\\
\exp(-x^2\sqrt{\alpha/D}/2) & ~~{\rm
for}~~|x| \to \infty.\\
               \end{array}
        \right. 
\l{eq:stationary-px-parabolic-resetting-limiting-forms}
\eea

Using Eq. (\ref{eq:stationary-px-parabolic-resetting}) and the result ${\rm d}K_{1/4}(x)/{\rm
d}x=-(1/2)\left(K_{3/4}(x)+K_{5/4}(x)\right)$, it may be easily shown
that as $x \to 0^{\pm}$, one has ${\rm d}P_{\rm st}(x|x_0=0))/{\rm
d}x=\mp \sqrt{\alpha/D}\Gamma(3/4)/(\sqrt{\pi}\Gamma(1/4))$, implying
thereby that the first derivative of $P_{\rm st}(x|x_0=0)$ is discontinuous at $x=0$. We thus
conclude that the spatial distribution $P_{\rm st}(x|x_0=0)$ exhibits a cusp singularity at $x=0$. This feature of cusp singularity at the
resetting location $x_0=0$ is also seen in the stationary
distribution (\ref{eq:Pxstat-constant-resetting-final}), and is a
signature of the steady state being a nonequilibrium one~\cite{Evans:2011-1,Gupta:2014,Nagar:2016,Gupta:2016}. Note the existence of faster-than-exponential tails suggested by Eq.
(\ref{eq:stationary-px-parabolic-resetting-limiting-forms}) in comparison
to the exponential tails observed in the case of resetting at a constant
rate, see Eq. (\ref{eq:Pxstat-constant-resetting-final}). This is 
consistent with the fact that with respect to the case of resetting at a
space-independent rate, a parabolic
rate of resetting implies that the further the particle is from $x_0=0$, the more
enhanced is the probability that a resetting event takes place, and, hence, a smaller
probability of finding the particle far away from the resetting location.

Let us consider the case of an overdamped Brownian
particle that is trapped in a harmonic
potential $V(x) = (1/2)\kappa x^2$, with $\kappa>0$, and is undergoing the
Langevin dynamics (\ref{eq:eom}). At equilibrium, the distribution of the position of the particle is given by the Boltzmann-Gibbs distribution
\bea
P_{\rm eq} (x) &=&\exp(-\kappa x^2/2k_{\rm B}T)/Z,  
\l{eq:stationary-px-parabolic-resetting-limiting-forms-equivalence}
\eea
with $Z= \sqrt{2\pi k_{\rm B}T/\kappa}$ being the partition
function. Comparing
Eqs.~(\ref{eq:stationary-px-parabolic-resetting-limiting-forms})
and~(\ref{eq:stationary-px-parabolic-resetting-limiting-forms-equivalence}),
we see that using a harmonic potential with a suitable $\kappa$, the
stationary distribution of a free Brownian particle undergoing parabolic
resetting may be made to match in the tails with the stationary
distribution of a Brownian particle trapped in the 
harmonic potential and evolving in the absence of any resetting.
On the other hand, the cusp singularity in the former cannot be
achieved with the Langevin dynamics in any harmonic
potential without the inclusion of resetting events.

Let us note that the stationary states
(\ref{eq:Pxstat-constant-resetting-final}) and
(\ref{eq:stationary-px-parabolic-resetting}) are entirely induced by the
dynamics of resetting. Indeed, in the absence of any resetting, the dynamics of a
free diffusing particle does not allow for a long-time stationary state,
since in the absence of a force, there is no way in which the motion of
the particle can be bounded in space.
On the other hand, in presence of resetting, the dynamics of repeated
relocation to a given
position in space can effectively compete with the inherent tendency of the
particle to spread out in space, leading to a bounded motion, and, hence,
a relaxation to a stationary spatial distribution at long times.
In the next section, we consider the situation where the particle even
in the absence of any resetting has a localized stationary spatial distribution,
and investigate the change in the nature of the spatial distribution of
the particle owing to the inclusion of resetting events.

%%%%%%%%%%%%%%%%%%%%%%%%%%%%%%%%%%%%%%%%%%%%%%%%%%%%%%%%%%%%%%%%%%%%%%%%%%%%%%%%
%%%%%%%%%%%%%%%%%%%%%%%%%%%%%%%%%%%%%%%%%%%%%%%%%%%%%%%%%%%%%%%%%%%%%%%%%%%%%%%%
%%%%%%%%%%%%%%%%%%%%%%%%%%%%%%%%%%%%%%%%%%%%%%%%%%%%%%%%%%%%%%%%%%%%%%%%%%%%%%%%
\subsection{Particle trapped in a harmonic potential with energy-dependent resetting}
\l{subsec:parabolic-resetting-force}

\begin{figure}[!ht]
\centering
\includegraphics[width=0.45\textwidth]{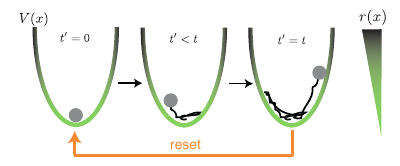}
\caption{\textbf{Illustration of the energy-dependent resetting of a Brownian particle
moving in a harmonic potential.}  A Brownian particle (grey circle) immersed in a
thermal bath at temperature $T$ moves with diffusion coefficient $D$,
with its motion being confined by a harmonic potential $V(x)=\kappa
x^2/2$ (green), with $\kappa$ being the stiffness constant. Here, $x$ is the position of the
particle with respect to the center of the potential. The particle,
initially located in the trap center ($t'=0$, left panel), diffuses at
subsequent times in the energy landscape ($t'<t$, middle panel), until a
resetting event occurs at time $t'=t$ (right panel). The black curve
represents the history of the particle from $t'=0$ up to the time corresponding to each snapshot. 
 The rate of resetting (right colorbar) is proportional to the instantaneous energy of the particle, and, therefore, a reset is more likely to take place as the particle climbs up the potential.}
\l{fig:energy-resetting}
\end{figure}

We now introduce a resetting problem that is relevant in physics: an
overdamped Brownian particle immersed in a thermal bath at temperature
$T$ and trapped with a harmonic potential centered at the origin:
$V(x)=(1/2)\kappa x^2$, where $\kappa>0$ is the stiffness constant of
the harmonic potential. The particle, initially located at
$x_0=0$, may be reset at any time $t$ to the origin with a
probability that depends on the energy of the particle at time $t$. The
dynamics is shown schematically in Fig. \ref{fig:energy-resetting}. For
purposes of illustration of the nontrivial effects of resetting, we
consider the following space-dependent reseting rate:
\be
r(x)=\frac{3}{2\tau_c}\frac{V(x)}{k_{\rm B}T} =\frac{3}{4}
\frac{\mu^2\kappa^2}{D}x^2,
\l{eq:re1}
\ee
where we use $D=k_BT \mu$ in obtaining the second equality. Note that the resetting rate is proportional to the energy of the
particle (in units of $k_{\rm B}T$) divided by the timescale $\tau_c
\equiv 1/\mu\kappa$ that characterizes the relaxation of the particle in
the harmonic potential in the absence of any resetting. In this way, it
is ensured that the rate of resetting~(\ref{eq:re1}) has units of
inverse time. Note also that
in the absence of any resetting, the particle relaxes to an equilibrium
stationary state with a spatial distribution given by the usual
Boltzmann-Gibbs form: 
\be
P_{\rm st}^{r(x)=0}(x)=\sqrt{\frac{\kappa}{2\pi k_BT}}\exp\left(-\frac{\kappa
x^2}{2k_BT}\right).
\l{eq:BGform-parabola-parabola}
\ee

Using $F(x) = -\partial_x V(x)=-\kappa x$ and the
expression~(\ref{eq:re1}) for the resetting rate in
Eq.~(\ref{eq:quantum-potential}), we find that the potential of the
corresponding quantum mechanical problem is given by
\be
V_{\rm q}(x)=\frac{\mu^2(F(x))^2}{4D}+\frac{\mu F'(x)}{2}+r(x) =  \frac{\mu^2\kappa^2 x^2}{D}-\frac{\mu \kappa}{2}, 
\l{eq:Vq-parabola-parabola}
\ee
where we have used $F'(x)=-\kappa$. From Eqs.~(\ref{eq:pnoresq})
and~(\ref{eq:qma}), we obtain
\bea
&&P_{\rm no\;res}(x,t|x_0=0,0)=\exp\left(\frac{\mu}{2D} \int_{x_0}^x F(x)~{\rm
d}x\right)\nonumber\\
&&\times \exp\left(\frac{\mu\kappa t}{2}\right)\langle
x|\exp(-H_{\rm q}t)|x_0=0\rangle\nonumber\\
&&=\exp\left(-\frac{x^2}{4D\tau_c}\right)\exp\left(\frac{t}{2\tau_c}\right)\langle
x|\exp(-H_{\rm q}t)|x_0=0\rangle,\nonumber\\
\l{eq:Pnoreset-parabola-parabola-0}
\eea
where the quantum Hamiltonian is given by
\be
H_{\rm q}= -\frac{1}{2m_{\rm q}}\frac{\partial^{2}}{\partial
x^{2}} + \frac{\mu^2\kappa^2 x^2}{D};~~m_{\rm q}=\frac{1}{2D},~\hbar=1. 
\l{eq:Hq-parabola-parabola}
\ee
We thus find that the propagator $\langle x|\exp(-H_{\rm
q}t)|x_0=0\rangle$ is given by the propagator of a quantum harmonic
oscillator, which has been calculated in
Sec.~\ref{subsec:parabolic-resetting}. In fact, the Hamiltonian given by
Eq.~(\ref{eq:Hq-parabola-parabola}) is identical to that in
Eq.~(\ref{eq:Hq-free-parabola}) with the identification
$\alpha = \mu^2\kappa^2/D=1/D\tau_c^2$, so that by substituting $x_0=0$ and
$\alpha=1/(D\tau_c^2)$ in Eq.~(\ref{eq:parabolic-resetting-Pxt-x0}), we obtain
\bea
&&\langle x|\exp(-H_{\rm q}t)|x_0=0\rangle =\nonumber\\
&&\frac{1}{\sqrt{2\pi D\tau_c\sinh(2t/\tau_c)}}\exp\left(-\frac{x^2}{2D\tau_c}\coth(2t/\tau_c)\right). 
\l{eq:Pnoreset-parabola-parabola-1}\nonumber\\
\eea
From Eqs.~(\ref{eq:Pnoreset-parabola-parabola-0}) and (\ref{eq:Pnoreset-parabola-parabola-1}), we obtain
\bea
&& P_{\rm no\;res}(x,t|x_0=0,0)=\nonumber\\
&&\frac{\exp(t/2\tau_c)}{\sqrt{2\pi D\tau_c\sinh(2t/\tau_c)}}\exp\left(-\frac{x^2}{2D\tau_c}\left[ \frac{1}{2}+\coth(2t/\tau_c)\right]\right).\nonumber\\
\l{eq:Pnoreset-parabola-parabola-final}
\eea

Following Eq.~(\ref{eq:Pt-final}), we may now calculate the probability
of the first-reset time by using
Eq.~(\ref{eq:Pnoreset-parabola-parabola-final}) to get
\bea
&& P_{\rm res}(t|x_0=0)=\frac{\exp(t/2\tau_c)}{\sqrt{2\pi D\tau_c\sinh(2t/\tau_c)}}\nonumber\\
&&\times\int_{-\infty}^{\infty}{\rm d}x\frac{3x^2}{4D\tau_c^2}  \exp\left(-\frac{x^2}{2D\tau_c}\left[ \frac{1}{2}+\coth(2t/\tau_c)\right]\right)\nonumber\\ 
&&=\frac{3\exp(t/2\tau_c)}{4\tau_c}\sqrt{\frac{1}{\sinh(2t/\tau_c)(1/2+\coth(2t/\tau_c))^3}},
\l{eq:Preset-parabola-parabola-final}\nonumber\\
\eea
which may be checked to be normalized: $\int_0^{\infty} {\rm d}t~P_{\rm res}(t|x_0=0) =
1$. The first-reset time
distribution~(\ref{eq:Preset-parabola-parabola-final}) may be written in
the scaling form
\bea
P_{\rm res}(t|x_0=0)&=&\frac{3}{4\tau_c}{\cal G}\!\left(\frac{2t}{\tau_c} \right),
\l{eq:Preset-parabola-parabola-final2}
\eea
with the scaling function given by ${\cal
G}(y)=\exp(y/4)\sinh(y)^{-1/2}(1/2+\coth(y))^{-3/2}$. 

\begin{figure}[ht]
\centering
\includegraphics[width=0.4\textwidth]{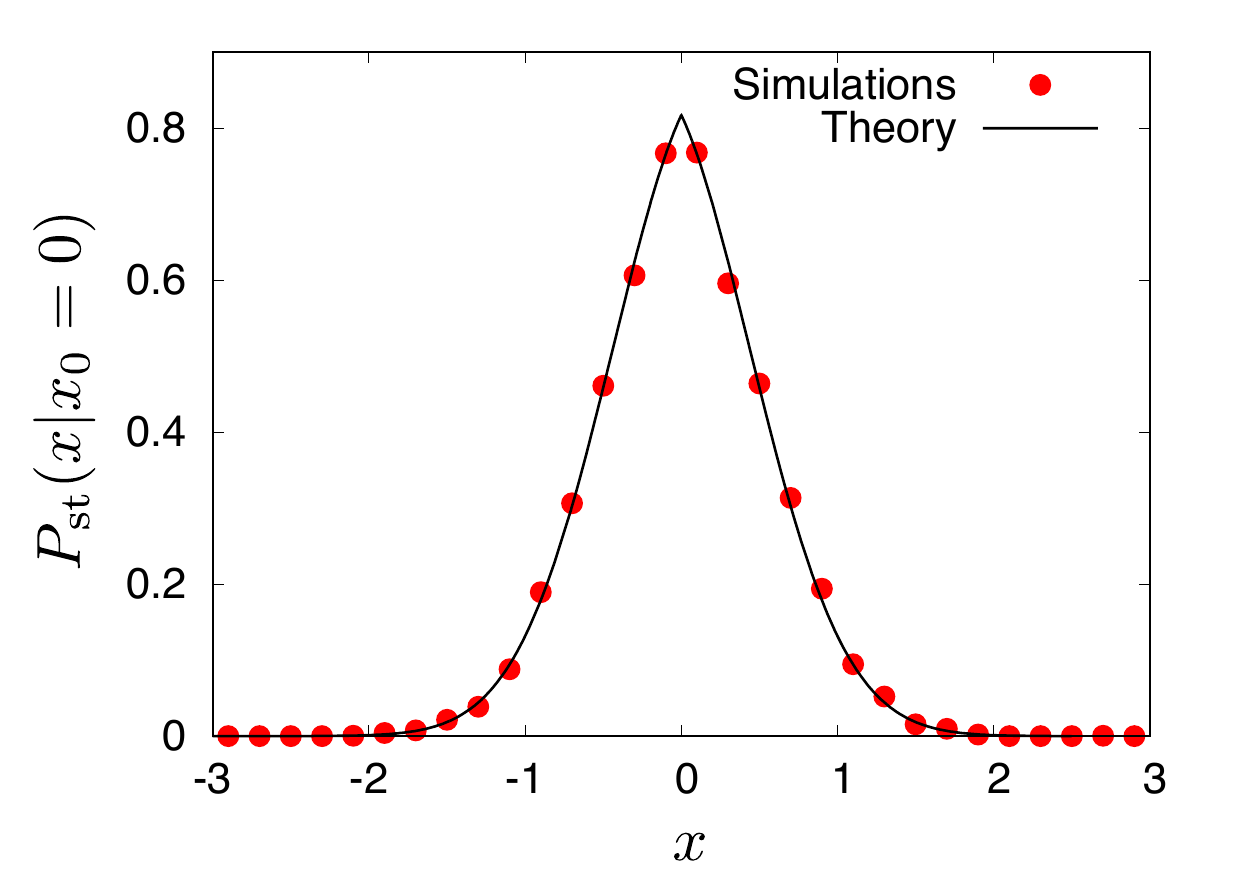}
\caption{\textbf{Theory versus simulation results for the stationary
spatial distribution of a Brownian particle trapped in a harmonic
potential and undergoing energy-dependent resetting.} The points denote simulation results, while the line stands
for the exact analytical expression given in Eq.
(\ref{eq:stationary-px-parabola-parabolic-resetting}). The numerical results were obtained from 
$10^4$ independent simulations of the Langevin dynamics described in Section
\ref{sec:model}.  The
error bars associated with the data points are smaller than the symbol
size. The parameter values are
$D=1.0,\tau_c=0.5$.}
\label{fig:parabola-parabola}
\end{figure}
The mean first-reset time, given by $\langle t\rangle_{\rm res}\equiv\int_0^{\infty} {\rm d}t P_{\rm res}(t|x_0=0)$,
equals 
\bea
&&\langle t \rangle_{\rm res} =
\frac{4\tau_c}{\sqrt{3}}\,
_2F_1\!\left(\frac{1}{8},\frac{1}{2};\frac{9}{8};-\frac{1}{3}\right),
\l{eq:treset-parabola-parabola}
\eea
where $_pF_q(a_1,a_2,\ldots,a_p;b_1,b_2,\ldots,b_q;x)$ is the
generalized hypergeometric function. Introducing the variable $z\equiv
2t/\tau_c$, and using Eq. (\ref{eq:Pnoreset-parabola-parabola-final}), we get
\bea
&&P_{\rm st}(x|x_0=0)=\frac{1}{\langle t \rangle_{\rm res}
}\sqrt{\frac{\tau_c}{8\pi
D}}\exp\left(-\frac{x^2}{4D\tau_c}\right)\nonumber\\
&&\times \int_0^{\infty}{\rm d}z
\frac{\exp(z/4)}{\sqrt{\sinh(z)}}\exp\left(-\frac{x^2\coth z}{2D\tau_c}\right)\nonumber \\
&&=\frac{1}{\langle t \rangle_{\rm res} }\sqrt{\frac{\tau_c}{8\pi D}}\exp\left(-\frac{x^2}{4D\tau_c}\right)\nonumber\\
&&\times \frac{1}{2}\left( \frac{x^2}{4D\tau_c}  \right)^{-1/4}\Gamma(1/8)\,W_{1/8,1/4}\left( \frac{x^2}{D\tau_c}\right), 
\l{eq:Pst-parabola-parabola}
\eea
where $W_{\mu,\nu}$ is Whittaker's W function.

Using
Eq.~(\ref{eq:treset-parabola-parabola}) in Eq.~(\ref{eq:Pst-parabola-parabola}), we obtain
\bea
&&P_{\rm st}(x|x_0=0)=\frac{(1/8)\Gamma(1/8)}{\,_2F_1\!\left(\frac{1}{8},\frac{1}{2};\frac{9}{8};-\frac{1}{3}\right)}\nonumber\\
&&\times  \sqrt{\frac{3}{8\pi D\tau_c}}\exp\left(-\frac{x^2}{4D\tau_c}\right)\left( \frac{4D\tau_c}{x^2}  \right)^{1/4}\,W_{1/8,1/4}\left( \frac{x^2}{D\tau_c}\right), \nonumber\\
\label{eq:stationary-px-parabola-parabolic-resetting}
\eea
which may be checked to be normalized to unity. We may write the stationary distribution
in terms of a scaled position variable as
\bea
P_{\rm st}(x|x_0=0)= \frac{(1/8)\Gamma(1/8)}{\,_2F_1\!\left(\frac{1}{8},\frac{1}{2};\frac{9}{8};-\frac{1}{3}\right)}\sqrt{\frac{3}{8\pi D\tau_c}}\mathcal{R}\!\left( \frac{x}{\sqrt{2D\tau_c}}  \right),\nonumber\\
\l{eq:statparresparpot}
\eea
with $\mathcal{R}(y) \equiv
\exp(-y^2/2)(2/y^2)^{1/4}\,W_{1/8,1/4}(2y^2)$ being the scaling
function.
 The expression
(\ref{eq:stationary-px-parabola-parabolic-resetting}) is checked in simulations in Fig.
\ref{fig:parabola-parabola}. The simulations involved
numerically integrating the dynamics described in Section
\ref{sec:model}, with integration timestep equal to $0.01$. 
Using the results that as $x\to 0$, we have
$W_{k,\mu}(x)=\frac{\Gamma(2\mu)}{\Gamma(1/2+\mu-k)}x^{1/2-\mu}$
for $0\le {\rm Re}~\mu<1/2,~\mu \ne 0$ and that as $x\to \infty$, we have
$W_{k,\mu}(x)\sim e^{-x/2}x^{k}$ for real $x$
\cite{Olver:2016}, we get
\bea
P_{\rm st}(x|x_0=0) \sim
\left\{
\begin{array}{ll}
\frac{(1/8)\Gamma(1/8)}{\thinspace_{2}F_{1}(1/8,1/2;9/8;-1/3)}\thinspace\frac{\sqrt{3/8D\tau_{c}}}{\Gamma(5/8)} & {\rm for}~x\to 0,\\
\exp\Big(-\frac{3x^{2}}{4D\tau_{c}}\Big) & {\rm
for}~|x| \to \infty.\nonumber\\
               \end{array}
        \right. 
\l{eq:stationary-px-parabola-parabola-resetting-limiting-forms}
\eea
We see again the existence of a cusp at the resetting location $x_0=0$,
similar to all other cases we studied in this paper.

The results of this subsection could inspire future experimental studies
using optical tweezers in which the resetting protocol could be
effectively implemented by using feedback control~\cite{Berut:2012,Roldan:2014,Koski:2014}. Interestingly,
colloidal and molecular gelly and glassy systems show hopping motion
of their constituent particles between potential traps or ``cages," the
latter originating from the interaction of the particles with their
neighbors~\cite{Sandalo:2017}. Such a phenomenon is also exhibited by
out-of-equilibrium glasses and gels during the process of
aging~\cite{Ludovic:2011}. Our results in this section could provide valuable insights
into the aforementioned dynamics, since the emergent potential cages may
be well approximated by harmonic traps and the hopping process as a resetting event.

\section{Shortcuts to confinement}

A hallmark of the examples solved exactly in Sec.~\ref{sec:applications}
by using our path-integral formalism is the existence of stationary
distributions with prominent {\em cusp singularities} (see
Figs.~\ref{fig:free-parabola} and~\ref{fig:parabola-parabola}). These
examples demonstrate that the particle can be confined around a
prescribed location by using appropriate space-dependent rates of resetting.  

In physics and nanotechnology, the issue of achieving an accurate control of fluctuations of
small-sized particles is nowadays attracting considerable
attention~\cite{Martinez:2013,Berut:2014,Dieterich:2015}. For instance,
using optical tweezers and noisy electrostatic fields, it is now
possible to control accurately the amplitude of fluctuations of the
position of a Brownian
particle~\cite{Martinez:2017,Gavrilov:2017,Ciliberto:2017}. Such
fluctuations may be characterized by an effective temperature. Experiments have reported effective temperatures of a colloidal particle
in water up to 3000K~\cite{Martinez:2013}, and have recently been used to
design colloidal heat engines at the mesoscopic
scale~\cite{Martinez:2016,Martinez:2017}. Effective
confinement of small systems is of paramount importance for success of
quantum-based computations with, e.g., cold atoms~\cite{Cirac:1995,Bloch:2005}. 

Does stochastic resetting provide an efficient way to reduce the
amplitude of fluctuations of a Brownian particle, thereby providing a
technique to reduce the associated effective temperature?
 We now provide some insights into this question.
 
Consider the following example of a nonequilibrium protocol: i) a
Brownian particle is initially confined  in a harmonic trap with a
potential $V(x)=(1/2)\kappa x^2$ for a sufficiently long time such that
it is in an equilibrium state with spatial distribution $P_{\rm
eq}(x)=\exp(-\kappa x^2/2k_{\rm B}T)/Z$, with $Z= \sqrt{2\pi k_{\rm
B}T/\kappa}$; ii) a space-dependent (parabolic) resetting rate
$r(x)=(3/2\tau_c)V(x)/k_{\rm B}T$, with $\tau_c=(1/\mu \kappa)$, is
suddenly switched on by an external agent. In order words, the rate of
resetting is instantaneously quenched from $r(x)=0$ to
$r(x)=(3/4)(\mu^2\kappa^2/D)x^2$; iii) the particle is let to
relax to a new stationary state in the presence of the trapping potential and parabolic resetting.  At the end of the protocol, the particle relaxes to the stationary distribution $P_{\rm st}(x)$ given by Eq.~(\ref{eq:statparresparpot}).

\begin{figure}[h]
\centering
\includegraphics[width=0.4\textwidth]{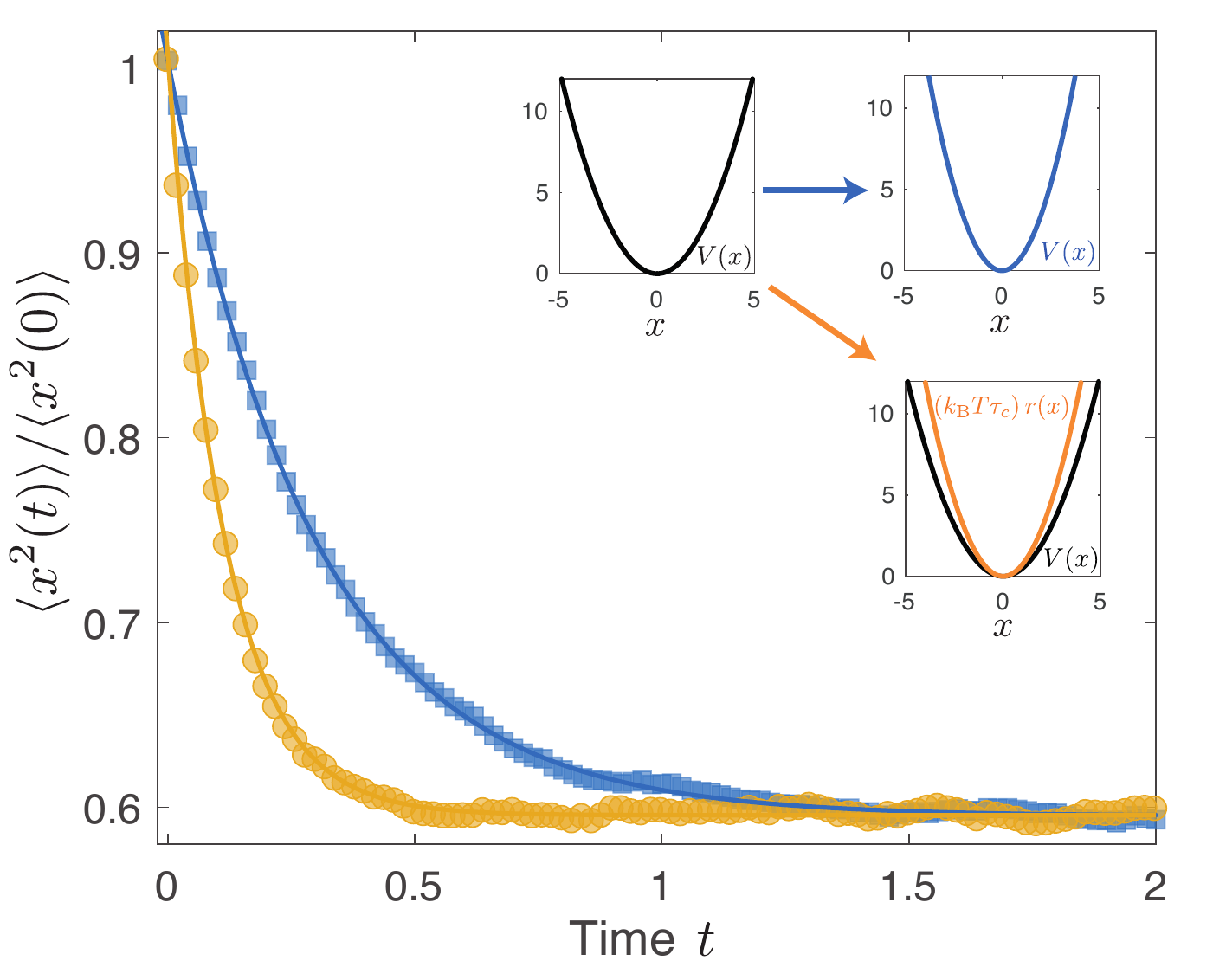}
\caption{\textbf{Shortcut to particle confinement.} Confinement of a
Brownian particle using harmonic potentials and space-dependent rates of
resetting. The symbols represent simulation results for relaxation
processes of $10^5$ non-interacting Brownian particles that are
initially in equilibrium in a harmonic potential $V(x)=(1/2)\kappa
x^2$. The blue circles represent the mean-squared displacement of the
particle in units of $\langle x^2(0)\rangle$ after an instantaneous
quench to a harmonic potential with stiffness $\kappa'=1.7\kappa$ (the blue
arrow in the inset). The yellow circles on the other hand represent the
mean-squared displacement of the particle in units of $\langle
x^2(0)\rangle$ after an instantaneous quench of the rate of resetting
from $r(x)=0$ to $r(x)=(3/4)(\mu^2\kappa^2/D)x^2$ (the orange arrow in
the inset). For the case in which the
 stiffness of the potential is quenched from $\kappa$ to $\kappa'$, it
 is easily seen that $\langle x^2(t)\rangle=\langle x^2 (0)\rangle
 \exp(-2\mu k' t)+(D/(\mu k'))[1-\exp(-2\mu k't)]$, thereby implying a
 relaxation timescale $\tau_{\rm quench}=1/(2\mu k')$ and yielding the corresponding curve in the figure. Note that
 the time in the $x$-axis is measured in units of $\tau_c= 1/\mu \kappa$. The other
 curve depicting the process of relaxation in presence of resetting may
 be fitted to a good approximation to $A+B e^{-t/\tau_{\rm reset}}$. One observes
 that $\tau_{\rm reset} \approx \tau_{\rm quench}/3$. 
The parameter values are $D=1$, $\kappa=1$ and $\mu=10$.}
\label{fig:quenchreset}
\end{figure}

We first note that before the sudden switching on of the resetting
dynamics, which we assume to happen at a reference time instant $t=0$, the
mean-squared displacement of the particle is given by
\be
\langle x^2(0)\rangle = \frac{k_{\rm B}T}{\kappa},
\ee
which follows from the equilibrium distribution before the resetting is
switched on, and is in agreement with
the equipartition theorem $\kappa \langle x^2(0)\rangle/2 =k_{\rm
B}T/2$. After the sudden switching on of the space-dependent rate of
resetting, the variance of the position of the particle relaxes at long
times to the stationary value
\be
\langle x^2(\infty)\rangle=\frac{k_{\rm B}T/\kappa}{\sqrt{3}\,
_2F_1\!\left(\frac{1}{8},\frac{1}{2};\frac{9}{8};-\frac{1}{3}\right)}\simeq
0.59 \frac{k_{\rm B}T}{\kappa},
\ee
as follows from Eq. (\ref{eq:statparresparpot}). 
The resetting dynamics induces in this case a reduction by about $40\%$ of the 
 variance of the position of the particle with respect to its initial
 value. We note that such a reduction of the amplitude of fluctuations
 of the particle could also have been achieved by performing a sudden
 quench of the stiffness of the harmonic potential by increasing its
 value from $\kappa$ to
 $\kappa'\simeq  (1/0.59) \kappa \simeq 1.7 \kappa$, without the need
 for switching on of resetting events.  To understand the difference
 between the two scenarios, it is instructive to compare the time
 evolution of the mean-squared displacement $\langle x^2(t)\rangle$
 towards the stationary value in the two cases, see inset in
 Fig.~\ref{fig:quenchreset}. We observe that resetting leads to the same
 degree of confinement in a shorter time. For the case in which the
 stiffness of the potential is quenched from $\kappa$ to $\kappa'$, it
 is easily seen that $\langle x^2(t)\rangle=\langle x^2 (0)\rangle
 \exp(-2\mu k' t)+(D/(\mu k'))[1-\exp(-2\mu k't)]$, thereby implying a
 relaxation timescale $\tau_{\rm quench}=1/(2\mu k')$ and yielding
 the corresponding curve in Fig. \ref{fig:quenchreset}. The other
 curve depicting the process of relaxation in presence of resetting may
 be fitted to a good approximation to $A+B e^{-t/\tau_{\rm reset}}$. One observes
 that $\tau_{\rm reset} \approx \tau_{\rm quench}/3$. 
Thus, for the example at hand, we may conclude that a sudden quench of
resetting profiles provides a {\em shortcut to confinement} of the position of
the particle to a desired degree with respect to a potential quench.
Similar conclusions were arrived at for mean first-passage times of resetting processes and equivalent equilibrium dynamics~\cite{Evans:2013}. 

It may be noted that the confinement protocol by a sudden quench of
resetting profiles introduced above is
amenable to experimental realization. Using microscopic particles
trapped with optical tweezers~\cite{Martinez:2017,Ciliberto:2017} or
feedback traps~\cite{Jun:2012,Gavrilov:2017,Gavrilov:2014}, it is now
possible to measure and control the position of a Brownian particle
with subnanometric precision. Recent experimental setups allow to exert
random forces to trapped particles, with a user-defined statistics for the random force~\cite{Martinez:2013,Berut:2014,Martinez:2014,Martinez:2017}.  
The shortcut protocol using resetting could be explored in
the laboratory by designing a feedback-controlled experiment with optical
tweezers and by employing random-force generators according to, e.g., the protocol sketched in Fig.~\ref{fig:energy-resetting}.

%%%%%%%%%%%%%%%%%%%%%%%%%%%%%%%%%%%%%%%%%%%%%%%%%%%%%%%%%%%%%%%%%%%%%%%%%%%%%%%%
\section{Conclusions and outlook}
\l{sec:conclusion}
In this paper, we addressed the fundamental question of what happens when a continuously evolving stochastic process is repeatedly interrupted at
random times by a sudden reset of the state of the process to a given fixed state. To this end,
we studied the dynamics of an overdamped Brownian particle diffusing in force fields and resetting to a given spatial location with a rate that has an essential dependence on space, namely,
the probability with which the particle resets is a function of the current location of the particle. 

To address stochastic resetting in the aforementioned scenario, we
employed a path-integral approach, discussed in detail in
Eqs.~(\ref{eq:pnoresq})-(\ref{eq:quantum-potential}) in
Sec.~\ref{sec:propagator}.  Invoking the Feynman-Kac formalism, 
we obtained an equality that
relates the probability of transition between different spatial
locations of
the particle before it encounters any reset to the quantum propagator of
a suitable quantum mechanical problem (see Sec.~\ref{sec:propagator}).
Using this formalism and elements from renewal theory, we obtained
closed-form analytical expressions for a number of statistics of the dynamics, e.g., the probability distribution of the first-reset time (Sec.~\ref{sec:prob-first-reset}), the time-dependent spatial distribution (Sec.~\ref{sec:prob-distr}), and the stationary spatial distribution (Sec.~\ref{sec:stat-distr}). 

We applied the method to a number of representative examples, including in particular those involving 
nontrivial spatial dependence of the rate of resetting.  Remarkably, we
obtained the exact distributions of the aforementioned dynamical
quantifiers for two non-trivial problems: the resetting of a free
Brownian particle under ``parabolic'' resetting
(Sec.~\ref{subsec:parabolic-resetting}) and the resetting of a Brownian
particle moving in a harmonic potential with a resetting rate that
depends on the energy of the particle~(Sec.
\ref{subsec:parabolic-resetting-force}). For the latter case, we showed that using instantaneous quenching of resetting profiles allows to restrict the mean-squared
displacement of a Brownian particle to a desired value on a faster
timescale than by using instantaneous potential quenches. We expect that
such a {\em shortcut to
confinement} would provide novel insights in ongoing research on,
e.g., engineered-swift-equilibration protocols~\cite{ESE,Granger:2016} and shortcuts to adiabaticity~\cite{Deffner:2013,Deng:2013,Tu:2014}.

Our work may also be extended to treat systems
of interacting particles, with the advantage that the corresponding
quantum mechanical system can be treated effectively by using tools of
quantum physics and many-body quantum theory. Our approach also provides a
viable method to calculate path probabilities of complex stochastic processes. Such calculations are of particular interest in many contexts, e.g., in stochastic
thermodynamics~\cite{Jarzynski:2011,Seifert:2012,Celani:2012,Bo:2017}, and in the study of several biological systems such as molecular
motors~\cite{Julicher:1997,Guerin:2011}, active gels~\cite{Basu:2008}, genetic switches~\cite{Perez-Carrasco:2016,Schultz:2008}, etc.  As a
specific application in this direction, our approach allows to explore
the physics of~{\em Brownian tunnelling}~\cite{Roldan}, an interesting
stochastic resetting version of the well-known phenomenon of quantum
tunneling, which serves to unveil the subtle effects resulting from
stochastic resetting in, e.g., transport through nanopores~\cite{Trepagnier:2007}.

\section{Acknowledgements}
ER thanks Ana Lisica and Stephan Grill for initial discussions, Ken
Sekimoto and Luca Peliti for discussions on path integrals, and Domingo S\'anchez and Juan M. Torres for discussions on quantum mechanics.


\begin{thebibliography}{99}
\bibitem{Wick}{In quantum mechanics, this time transformation is often
called the Wick's rotation in honor of Gian-Carlo Wick}
\end{thebibliography}
\end{document}